\documentclass{elsart5}

 \usepackage{graphicx}

\usepackage{amssymb,amsbsy,amsmath}

\newcommand{\op}[1]{%
    \fontdimen12\textfont3=2pt\fontdimen12\scriptfont3=1.4pt%
    \!\null\mathop{\vphantom{#1}\smash{#1}}\limits_{\sim}\null\!}
\newcommand{\fmref}[1]{(\protect\ref{#1})}

\begin{document}

\begin{frontmatter}

\title{Revisiting and modeling the magnetism of hole-doped
  CuO$_2$ spin chains in Sr$_{14-x}$Ca$_x$Cu$_{24}$O$_{41}$}

\author[aff1]{J. Schnack\corauthref{cor1}}
\ead{jschnack@uos.de}
\corauth[cor1]{}
\author[aff2]{R. Klingeler}
\author[aff2]{V. Kataev}
\author[aff2]{B. B\"uchner}
\address[aff1]{University of Osnabr\"uck, Dept. of Physics, D-49069 Osnabr\"uck, Germany}
\address[aff2]{Institute for Solid State and Materials Research IFW Dresden, P.O. Box
270116, D-01171 Dresden, Germany}
\received{12 June 2005}
\revised{13 June 2005}
\accepted{14 June 2005}


\begin{abstract}
  Magnetization measurements of
  Sr$_{14-x}$Ca$_x$Cu$_{24}$O$_{41}$ with $0 \le x \le 12$ in
  magnetic fields up to 16~T show that the low-temperature
  magnetic response of the CuO$_2$ spin chains changes strongly
  upon Ca doping. For $x=0$ quantum statistical simulations
  yield that the temperature and field dependence of the
  magnetization can be well described by an effective Heisenberg
  model in which the ground state configuration is composed of
  spin dimers, trimers, and monomers. For $x>0$ a constant contribution
  to the low-temperature magnetic susceptibility is observed
  which cannot be explained in terms of simple chain models.
  Alternative scenarios are outlined.
\end{abstract}


\end{frontmatter}

\section{Sr$_{14}$Cu$_{24}$O$_{41}$}
\label{sec-1}

\begin{figure}[ht!]
\begin{center}
\includegraphics[clip,width=70mm]{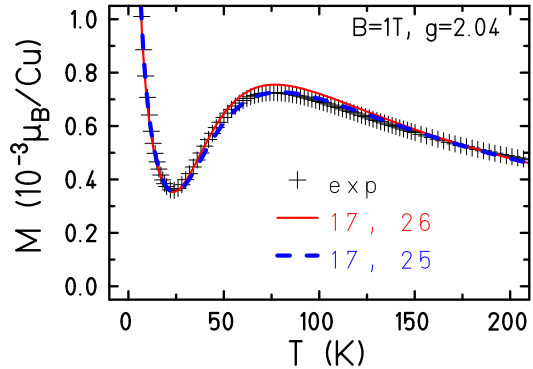}
\caption{Magnetization versus temperature for
  $B=1$~T: Experimental data (B parallel to the $c$-axis) are
  given by crosses. The results of a complete numerical
  diagonalization are depicted by a solid curve for
  $N_{\text{s}}=17$ and $N_{\text{h}}=26$ as well as by a dashed
  curve for $N_{\text{s}}=17$ and $N_{\text{h}}=25$.}
\label{fig-1}
\end{center}
\end{figure}

The low-temperature magnetic response of
Sr$_{14}$Cu$_{24}$O$_{41}$ is dominated by the CuO$_2$ chain
subsystem, whereas the ladder subsystem is magnetically silent
due to a large spin gap. The magnetic susceptibility, shown in
Fig.~\ref{fig-1}, exhibits two pronounced features, a clear
maximum at about 75~K and a Curie-like divergence at $T=0$. The
maximum has been attributed to the existence of
antiferromagnetically coupled dimers on the chains, which are
supposed to host nearly six Zhang-Rice singlets (``holes") per 10
Cu sites.  The $T=0$ divergence was attributed to free Cu spins
residing somewhere in the crystal structure. In a recent
investigation \cite{KBK:PRB06} it could be shown by means of EPR
measurements that these impurities are located on the chains
since they experience the same chemical environment as the Cu
ions bound in dimers. Taken this into account a quantum
statistical simulation -- as outlined in Ref.~\cite{Sch:EPJB05}
-- was performed in order to treat both dimers and monomers and
other possible substructures on a common footing.  The model is
similar to a simple Born-Oppenheimer description where the
electronic Hamiltonian (here spin Hamiltonian) depends
parametrically on the positions of the classical nuclei (here
hole positions).  Each configuration $\vec{c}$ of holes and
spins defines a Hilbert space which is orthogonal to all Hilbert
spaces arising from different configurations. The Hamilton
operator $\op{H}(\vec{c})$ of a certain configuration $\vec{c}$
is of Heisenberg type, i.~e.
\begin{eqnarray}
\label{JS-1-1} \op{H} &=& \sum_{\vec{c}}\; \left( \op{H}(\vec{c}) + V(\vec{c}) \right)
\\
\label{JS-1-2} \op{H}(\vec{c}) &=& - \sum_{u<v}\; J_{uv}(\vec{c})\; \op{\vec{s}}(u) \cdot
\op{\vec{s}}(v)
\\
\label{JS-1-3} V(\vec{c}) &=& \frac{e^2}{4\pi\epsilon_0\,\epsilon_r\,r_0} \frac{1}{2}\; \sum_{u\ne
v}\; \frac{1}{|u-v|} \ .
\end{eqnarray}
$J_{uv}(\vec{c})$ are the respective exchange parameters which
depend on the configuration of holes.  For the theoretical
results four exchange parameters are used.  The strongest and
antiferromagnetic exchange $J=-134$~K is across one hole. The
exchange across two holes $J_\parallel=15$~K is ferromagnetic
as is the exchange $J_{NN}=100$~K of neighboring spins. We also
apply a coupling across one spin which is weaker than $J$ and
taken to be $J_{s}=-34$~K. The following statements are robust
for reasonable variations of $J_\parallel$, $J_{NN}$, and
$J_{s}$. Periodic boundary conditions are applied for the
following calculations.

Two very good fits are shown in Fig.~\ref{fig-1} (lines)
together with the experimental data points (crosses). The fits
yield that it is equally well possible that the chains contain
monomers as well as that they contain trimers.  The first case
is very similar to the one found using density functional theory
calculations \cite{GeL:PRL04,GeL:EPJB05B}.

\section{Sr$_{14-x}$Ca$_x$Cu$_{24}$O$_{41}$}
\label{sec-2}

\begin{figure}[ht!]
\begin{center}
\includegraphics[clip,width=70mm]{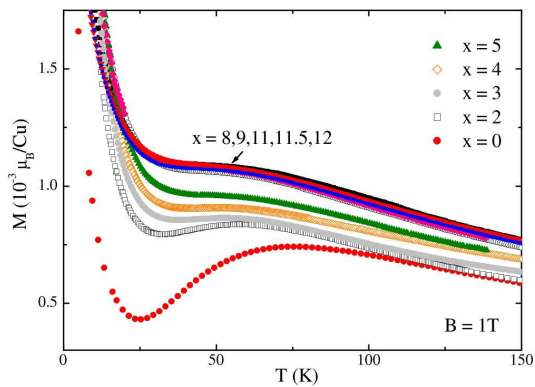}
\caption{Magnetization of Sr$_{14-x}$Ca$_x$Cu$_{24}$O$_{41}$
  with $0\leq x\leq 12$ in a magnetic field of $B=1$\,T parallel
  to the $c$-axis.}
\label{fig-2}
\end{center}
\end{figure}

For the calcium doped compounds
Sr$_{14-x}$Ca$_x$Cu$_{24}$O$_{41}$ the picture is much more
involved. Whereas Sr$_{14}$Cu$_{24}$O$_{41}$ exhibits charge
order on the chains due to the combined presence of long range
Coulomb interaction like in Na$_{1+x}$CuO$_{2}$ \cite{HSM:PRL05}
and commensurate doping level, the Ca-doping results in a
decreasing stability of charge order
\cite{ABC:PRB00,KCG:PRB01,HEB:PRL04}. In addition, the dimer
spin gap is strongly affected by the doping level, and at low
temperatures a finite susceptibility, which is not Curie-like,
is observed \cite{KTB:PRB05}. This susceptibility increases with
the Ca content. Figure~\ref{fig-2} shows the temperature
dependence of the magnetization for various Ca-doped compounds
for $B=1$~T.

It is known that application of chemical pressure through
Ca-doping leads to a hole transfer from the chains to the
ladders in Sr$_{14-x}$Ca$_x$Cu$_{24}$O$_{41}$
\cite{NMK:PRB00,KCG:PRB01}. The estimated maximal
reduction of the number of holes on the chains is one hole out
of six per formula unit. In addition, it is usually assumed that
the increased presence of holes on the ladders does not
significantly reduce the ladder's spin gap. Therefore, the
low-temperature low-field magnetic response should still origin
from the chain subsystem only.

\begin{figure}[ht!]
\begin{center}
\includegraphics[clip,width=70mm]{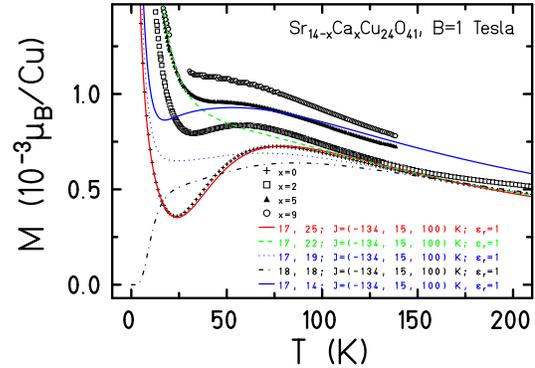}
\caption{Magnetization of model spin-hole chains for various
  numbers of spins and holes (curves). Experimental data are
  given by symbols.}
\label{fig-3}
\end{center}
\end{figure}

In the context of the quantum statistical model,
Eqs.~\fmref{JS-1-1}-\fmref{JS-1-3}, we investigate a huge
variety of doping levels numerically.  The number of spins is
$N_{\text{s}}=16, 17$, or $18$, whereas the number of holes is
varied in the range $26\geq N_{\text{h}}\geq 10$.
Figure~\ref{fig-3} shows a small selection of resulting
magnetization curves. A striking feature of all curves -- not
only those which are shown -- is that for $N_{\text{h}}\geq
N_{\text{s}}$ the high temperature part at $T=100\dots 150$~K
does not increase with decreasing number of holes. Only for hole
levels significantly less than 50~\% the chains show a
magnetization comparable to the experiment. 

The discrepancy between experimental and theoretical results has
the following consequences. Antiferromagnetic spin chains of
finite length (50~\% holes) cannot account for the observed
behavior. In general, hole levels above 50~\% cannot produce the
observed high-temperature magnetization. In addition both
scenarios cannot explain the low-temperature high-field
magnetization. Thermal and field induced dynamics of the holes
might be a solution. In addition, the assumption of a gapped
spectrum of hole-doped ladders needs to be justified. On the
theory side the assumption of hole localization should be
relaxed, i.e. the system should be investigated in the context
of a generalized Hubbard model.


\end{document}